# A MAPPING METHOD FOR THE GRAVITATIONAL FEW-BODY PROBLEM WITH DISSIPATION

RENU MALHOTRA

*Lunar and Planetary Institute, 3600 Bay Area Blvd, Houston, TX 77058*

*E-mail: renu@lpis39.jsc.nasa.gov*

**Abstract.** Recently a new class of numerical integration methods – "mixed variable symplectic integrators" – has been introduced for studying long-term evolution in the conservative gravitational few-body problem. These integrators are an order of magnitude faster than conventional ODE integration methods. Here we present a simple modification of this method to include small non-gravitational forces. The new scheme provides a similar advantage of computational speed for a larger class of problems in Solar System dynamics.

**Key words:** numerical integration, symplectic integrators, solar system dynamics

## 1. INTRODUCTION

Efforts to solve the question of the long-term stability of planetary orbits have recently spun-off a new algorithm (Wisdom & Holman 1991, Kinoshita *et al.* 1991) for the numerical integration of the conservative gravitational few-body problem described by a Hamiltonian of the following type:

$$H(\mathbf{r}_j, \mathbf{p}_j) = H_{\text{kepler}}(\mathbf{r}_j, \mathbf{p}_j) + \mu H'(\mathbf{r}_j). \tag{1}$$

Here $H_{\text{kepler}}(\mathbf{r}_j, \mathbf{p}_j)$ is the usual Hamiltonian for the two-body problem which describes the unperturbed Keplerian orbit of the $j$-th planet, and $\mu H'(\mathbf{r}_j)$ describes the much smaller mutual gravitational interactions of the planets. The new technique exploits two features of the system: (i) the symplectic structure of the equations of motion, and (ii) the integrability of both the unperturbed Hamiltonian, $H_{\text{kepler}}$, and the perturbation Hamiltonian, $\mu H'$, when taken separately. The latter property permits an efficient algorithm based upon the use of two sets of canonical variables (one for $H_{\text{kepler}}$ and the other for $\mu H'$). Kinoshita *et al.* call this method a "modified symplectic integrator", while Wisdom & Holman call it an "N-body map". Saha & Tremaine (1992) have reviewed a class of symplectic integrators that are particularly suitable for Solar System dynamics; they refer to these by the more descriptive name, "mixed variable symplectic" [MVS] integrators.

The defining feature of the MVS integrators is that they preserve the symplecticity of the system and they do not produce a secular error in its first integrals. Therefore, quantities such as total energy and angular momentum (or the Jacobi integral for the restricted three-body problem) do not exhibit secular changes and the position error grows only linearly with time in a numerical solution obtained with these integrators. In contrast, conventional integrators such as the Runge-Kutta introduce a so-called "numerical dissipation" such that the numerical solution exhibits a secular change of total energy even in a conservative system, and position errors grow quadratically with time. For this simple reason, symplectic integrators have a significant advantage over conventional integrators for long integrations of a planetary system.

Since the first application by Wisdom & Holman (1991) to integrate the outer planets for a billion years (obtaining results that compared very well with the earlier results of the 'Digital Orrery' calculations of Sussman & Wisdom (1988)), this method has been used for several other problems in Solar System dynamics. Sussman & Wisdom (1992) followed all the planets for 100 million years; Saha (1992) adopted this method for a numerical investigation of the chaotic dynamics of asteroids near the 3:1 Kirkwood Gap; Levison & Duncan (1993) and Holman & Wisdom (1993) have used it to explore the long term stability of test particle orbits in the outer Solar System.

While the method is demonstrably very useful for a certain class of problems in Solar System dynamics, it is obvious that the Hamiltonian system of Eqn. 1 is a highly idealized description of an N-body system — that of point masses subject only to mutual gravitational interactions according to Newtonian gravity. Although the planetary system today resembles closely this idealized Hamiltonian system, this was not the case in the early stages of the Solar System. Even if one confines oneself to the present state, for the dynamical behavior over billion year timescales one may worry about additional physics (cf. Nobili *et al.* 1989) — the effects of higher order gravitational moments of the bodies, general relativistic effects, mass loss of the Sun by radiation and the solar wind, etc. In questions related to the early stages of the Solar System and planet formation, point mass interactions provide the dominant physics, but other dissipative processes such as inelastic collisions and gas drag are also important. Moreover, there is a large class of problems concerning the contemporary Solar System, such as the ring systems and satellite systems of the planets, in which forces other than point mass gravity are quite important even when relatively small. Here the quadrupole and higher order moments of the central planet, and collisional or tidal dissipation have important qualitative effects on the long term evolution. Yet another example is cometary orbital evolution where non-gravitational effects such as outgassing are small but of great interest. As a final example, we can mention the evolution of very small particles (dust grains) in the Solar System which are influenced significantly by radiation forces.

Of the examples cited above, higher order gravitational moments are straightforwardly incorporated in the 'mixed variable symplectic' schemes because (i) they do not change the symplectic nature of the system, and (ii) they still allow a separation of the Hamiltonian into an unperturbed (Keplerian) and a perturbation part, each of which is separately integrable. However, dissipative forces cannot be accommodated within the 'mixed variable symplectic' schemes, for, in general, they destroy the symplectic structure of the equations of motion. Because weak dissipation is of interest in many Solar System problems, and many of those problems are not amenable to analytical methods of study, this necessitates numerical studies involving long integration times. Therefore, computationally efficient methods of integration for a few-body 'planetary system' with dissipation are always desirable. This reason has motivated the present work.

In Section 2, we describe a straightforward modification of the N-body map that allows the inclusion of small non-Hamiltonian perturbations. The new method may be called a "generalized mixed variable mapping". (In the rest of this paper, we refer to this method simply as the "mapping" when there is no risk of ambiguity.) In Section 3 we describe an application of this method to the restricted 3-body problem with gas drag. We give detailed numerical error analysis of this problem, comparing the new method with the commonly used fourth order Runge-Kutta integrator for first-order ODE's (see, for example, Press *et al.* 1989). This study demonstrates the superior performance of the new method for planetary dynamics problems with weak dissipation. Section 4 is a summary.

## 2. GENERALIZED MIXED VARIABLE MAPPING

First, we give a brief description of the 'mixed variable symplectic' method in the implementation introduced by Wisdom & Holman (1991). Operationally, this technique may be summarized as follows. In order to integrate the system of Eqn. 1 from time $t_0$ to time $t_0 + \tau$ (where $\tau$ is a small increment, typically less than $1/10$ of the smallest orbital period):

— advance $\mathbf{r}, \mathbf{p}$ according to $H_{\text{kepler}}$ for time $\frac{1}{2}\tau$;
— apply a "kick" of magnitude $\Delta \mathbf{p} = -\tau \mu \left( \partial H'/\partial \mathbf{r} \right)$ to the momenta, $\mathbf{p}$, and
— advance the *new* positions and momenta according to $H_{\text{kepler}}$ for time $\frac{1}{2}\tau$ once again.

We have dropped the index $j$ for clarity. Note that the first and last sub-steps involve evolution on the unperturbed orbit (a Keplerian "drift"). This can be trivially accomplished in terms of the Keplerian orbital elements (or more generally, action-angle variables), while the second sub-step is trivial in cartesian variables. Hence the terminology, 'mixed variable'. As pointed out by Wisdom & Holman (1991), a convenient and efficient scheme to advance the positions and velocities, $\mathbf{r}$ and $\mathbf{v} = \mathbf{p}/m$, in-between the "kicks" (when the motion is strictly on a Keplerian ellipse) is encapsulated in Gauss' $f$ and $g$ functions (see Battin 1987, or Danby 1988):

$$\begin{aligned} \mathbf{r}(t) &= f(t, t_0)\mathbf{r}_0 + g(t, t_0)\mathbf{v}_0, \\ \mathbf{v}(t) &= \dot{f}(t, t_0)\mathbf{r}_0 + \dot{g}(t, t_0)\mathbf{v}_0. \end{aligned} \quad (2)$$

This solution is possible because $(\mathbf{r}, \mathbf{v})$ and $(\mathbf{r}_0, \mathbf{v}_0)$ are non-collinear vectors that lie in the same plane. The $f$ and $g$ functions require a determination of only a few of the osculating orbital elements from the initial conditions $(\mathbf{r}_0, \mathbf{v}_0)$. One can intuitively recognize the advantage of this method: the unperturbed Keplerian motion is not completely reinvented at each step, as is the case in conventional integrators.

The algorithm described above is for a second order integrator; the position error per time step is $\mathcal{O}(\mu\tau^3)$. (We follow the convention that an order $n$ integrator has truncation error $\mathcal{O}(\tau^{n+1})$ per time step). The method can be generalized to higher order, and certain other refinements can be made. These improvements as well as the numerical stability and truncation errors of the MVS integrators are analyzed in detail by Yoshida (1990), Wisdom & Holman (1992) and Saha & Tremaine (1992).

Having described the 'mixed variable symplectic' method, we now turn to a straightforward exploitation of this scheme for weak dissipative forces superposed on nearly-Keplerian motion. Consider the following equations of motion:

$$\dot{\mathbf{r}} = \frac{\partial H}{\partial \mathbf{p}},$$

$$\dot{\mathbf{p}} = -\frac{\partial H}{\partial \mathbf{r}} + \varepsilon\mathbf{F}, \qquad (3)$$

where $H$ is as given in Eqn. 1, and

$$\varepsilon\mathbf{F} = \varepsilon\mathbf{F}(t, \mathbf{r}, \mathbf{v}), \qquad (4)$$

describes a small, "smooth", local dissipative force.

The modification of the MVS method for the above system is as follows. In order to solve the initial value problem with $(\mathbf{r}, \mathbf{v}) = (\mathbf{r}_0, \mathbf{v}_0)$ at time $t_0$ to obtain $(\mathbf{r}(t), \mathbf{v}(t))$ at time $t = t_0 + \tau$, where $\tau$ is small relative to the unperturbed orbital period, we retain the separation of $H$ into $H_{\text{kepler}}$ and $\mu H'$, and also retain the "drift"–"kick"–"drift"... sequence in the integrator. The "kicks" to the momenta are the same as before in the conservative case (arising from the mutual gravitational interactions), but the "drift" part is no longer a purely Keplerian motion. That is, in-between the "kicks" we must now advance the positions and momenta according to $H_{\text{kepler}}$ *plus* the effects of the perturbation, $\varepsilon\mathbf{F}$. Thus, we need to solve the following equations of motion in the first and third sub-steps:

$$\dot{\mathbf{r}} = \mathbf{v}$$

$$\dot{\mathbf{v}} = -\frac{GM}{r^3}\mathbf{r} + \varepsilon\mathbf{F} \qquad (5)$$

where $-GM/r$ represents the unperturbed Newtonian potential for the $j$-th body. Accordingly, we seek a generalization of Eqn. 2 that accomplishes the "drift" between the "kicks" to some desired accuracy.

Towards this end, we note that for an arbitrary $\varepsilon \mathbf{F}$, the motion, although nearly Keplerian, need not remain in a plane. Therefore, we introduce a generalization of Eqns. 2 as follows:

$$\mathbf{r}(t) = \tilde{f}(t,t_0)\mathbf{r}_0 + \tilde{g}(t,t_0)\mathbf{v}_0 + \tilde{h}(t,t_0)(\mathbf{r}_0 \times \mathbf{v}_0),$$
$$\mathbf{v}(t) = \dot{\tilde{f}}(t,t_0)\mathbf{r}_0 + \dot{\tilde{g}}(t,t_0)\mathbf{v}_0 + \dot{\tilde{h}}(t,t_0)(\mathbf{r}_0 \times \mathbf{v}_0), \tag{6}$$

where the *modified* $\tilde{f}$ and $\tilde{g}$ functions and the new function, $\tilde{h}$, take account of the external dissipation. One can see directly that $\tilde{f} - f$, $\tilde{g} - g$ and $\tilde{h}$ are all $\mathcal{O}(\varepsilon)$. Below we derive expressions for these $\mathcal{O}(\varepsilon)$ corrections.

Without loss of generality, the vector $\mathbf{F}$ can be written as follows:

$$\mathbf{F}(t,\mathbf{r},\mathbf{v}) = F_1(t,\mathbf{r},\mathbf{v})\mathbf{r}_0 + F_2(t,\mathbf{r},\mathbf{v})\mathbf{v}_0 + F_3(t,\mathbf{r},\mathbf{v})(\mathbf{r}_0 \times \mathbf{v}_0), \tag{7}$$

where

$$\begin{aligned} F_1(t,\mathbf{r},\mathbf{v}) &= \frac{1}{r_0}\left[\frac{\hat{\mathbf{r}}_0 \cdot \mathbf{F} - (\hat{\mathbf{r}}_0 \cdot \hat{\mathbf{v}}_0)\hat{\mathbf{v}}_0 \cdot \mathbf{F}}{1 - (\hat{\mathbf{r}}_0 \cdot \hat{\mathbf{v}}_0)^2}\right], \\ F_2(t,\mathbf{r},\mathbf{v}) &= \frac{1}{v_0}\left[\frac{\hat{\mathbf{v}}_0 \cdot \mathbf{F} - (\hat{\mathbf{r}}_0 \cdot \hat{\mathbf{v}}_0)\hat{\mathbf{r}}_0 \cdot \mathbf{F}}{1 - (\hat{\mathbf{r}}_0 \cdot \hat{\mathbf{v}}_0)^2}\right], \\ F_3(t,\mathbf{r},\mathbf{v}) &= \frac{(\mathbf{r}_0 \times \mathbf{v}_0) \cdot \mathbf{F}}{|\mathbf{r}_0 \times \mathbf{v}_0|^2}, \end{aligned} \tag{8}$$

where $\hat{\mathbf{r}}_0$ and $\hat{\mathbf{v}}_0$ are the unit vectors along $\mathbf{r}_0$ and $\mathbf{v}_0$, respectively.

Consider the Taylor series,

$$\mathbf{r}(t_0 + \tau) = \mathbf{r}_0 + \tau\dot{\mathbf{r}}_0 + \frac{1}{2}\tau^2\ddot{\mathbf{r}}_0 + \frac{1}{6}\tau^3\dddot{\mathbf{r}}_0 + \cdots \tag{9}$$

Now,

$$\begin{aligned} \dot{\mathbf{r}}_0 &= \mathbf{v}_0, \\ \ddot{\mathbf{r}}_0 &= -\frac{GM}{r_0^3}\mathbf{r}_0 + \varepsilon\left[F_{10}\mathbf{r}_0 + F_{20}\mathbf{v}_0 + F_{30}(\mathbf{r}_0 \times \mathbf{v}_0)\right], \\ \dddot{\mathbf{r}}_0 &= -GM\left[\frac{\mathbf{v}_0}{r_0^3} - \frac{3(\mathbf{r}_0 \cdot \mathbf{v}_0)\mathbf{r}_0}{r_0^5}\right] + \varepsilon\left[\dot{F}_{10}\mathbf{r}_0 + \dot{F}_{20}\mathbf{v}_0 + \dot{F}_{30}(\mathbf{r}_0 \times \mathbf{v}_0)\right], \end{aligned} \tag{10}$$

where

$$\begin{aligned} F_{k0} &= F_k(t_0,\mathbf{r}_0,\mathbf{v}_0), \\ \dot{F}_{k0} &= \frac{\partial}{\partial t}F_k(t_0,\mathbf{r}_0,\mathbf{v}_0) + \dot{\mathbf{r}}_0 \cdot \frac{\partial}{\partial \mathbf{r}}F_k(t_0,\mathbf{r}_0,\mathbf{v}_0) + \ddot{\mathbf{r}}_0 \cdot \frac{\partial}{\partial \mathbf{v}}F_k(t_0,\mathbf{r}_0,\mathbf{v}_0). \end{aligned} \tag{11}$$

Substituting Eqn. 10 in Eqn. 9, we obtain

$$\begin{aligned}\mathbf{r}(t) = & \left(1 - \frac{1}{2}\tau^2 \frac{GM}{r_0^3} + \frac{1}{6}\tau^3 \frac{3GM\,\mathbf{r}_0 \cdot \mathbf{v}_0}{r_0^5} + \cdots\right)\mathbf{r}_0 + \left(\tau - \frac{1}{6}\tau^3 \frac{GM}{r_0^3} + \cdots\right)\mathbf{v}_0 \\ & + \varepsilon\bigg[\left(\frac{1}{2}\tau^2 F_{10} + \frac{1}{6}\tau^3 \dot{F}_{10} + \cdots\right)\mathbf{r}_0 + \left(\frac{1}{2}\tau^2 F_{20} + \frac{1}{6}\tau^3 \dot{F}_{20} + \cdots\right)\mathbf{v}_0 \\ & + \left(\frac{1}{2}\tau^2 F_{30} + \frac{1}{6}\tau^3 \dot{F}_{30} + \cdots\right)(\mathbf{r}_0 \times \mathbf{v}_0)\bigg]\end{aligned} \quad (12)$$

Comparing with Eqn. 6, it is easy to see that, as expected, the $\mathcal{O}(\varepsilon^0)$ coefficients of $\mathbf{r}_0$ and $\mathbf{v}_0$ are the Taylor series expansions of Gauss' $f$ and $g$ functions. And, finally, to leading order in $\tau$ we have the following expressions for the $\mathcal{O}(\varepsilon)$ corrections:

$$\begin{aligned}\tilde{f}(t_0+\tau, t_0) - f(t_0+\tau, t_0) &= \frac{1}{2}\tau^2 \varepsilon F_{10}, \\ \tilde{g}(t_0+\tau, t_0) - g(t_0+\tau, t_0) &= \frac{1}{2}\tau^2 \varepsilon F_{20}, \\ \tilde{h}(t_0+\tau, t_0) &= \frac{1}{2}\tau^2 \varepsilon F_{30}.\end{aligned} \quad (13)$$

Provided $\mathbf{F}(t, \mathbf{r}, \mathbf{v})$ is a sufficiently smooth function, the terms higher order in $\tau$ can be made negligibly small with suitably small step sizes.

Straightforward bookkeeping shows that the leading order truncation errors per time step with the above scheme are $\mathcal{O}(\mu\tau^3) + \mathcal{O}(\varepsilon\tau^3)$ in the position, and $\mathcal{O}(\mu\tau^3) + \mathcal{O}(\varepsilon\tau^2)$ in the velocity, where the first term derives from the gravitational perturbation $\mu H'$ and the second from the drag force. Since the position error per time step is $\mathcal{O}(\tau^3)$, the order of the integration is formally 2.

## 3. NUMERICAL RESULTS

In this section, we apply the new method to one of the simplest model problems in Solar System dynamics with dissipation. We give a detailed numerical error analysis, comparing the new method with the standard fourth order Runge-Kutta method (RK4).

The model problem we have chosen is the planar circular restricted three body problem with gas drag. The system consists of a test particle in a heliocentric orbit perturbed by a planet; the test particle orbit decays under the influence of a gas drag force whose magnitude is proportional to the square of the relative velocity of the particle with respect to the ambient gas. The gravitational perturbations on the particle are described by the Hamiltonian

$$H = \frac{1}{2}v^2 - \frac{GM_\odot}{r} - Gm_\mathrm{p}\left(\frac{1}{|\mathbf{r}-\mathbf{r}_\mathrm{p}|} - \frac{\mathbf{r}\cdot\mathbf{r}_\mathrm{p}}{r_\mathrm{p}^3}\right), \quad (14)$$

where $M_\odot$ is the mass of the Sun, and $\mathbf{r}, \mathbf{v}$ are the heliocentric position and velocity of the test particle; the heliocentric position of the planet, $\mathbf{r}_\mathrm{p}$ is given by

$$\mathbf{r}_\mathrm{p}(t) = a_\mathrm{p}(\cos n_\mathrm{p} t, \sin n_\mathrm{p} t), \qquad (15)$$

and $m_\mathrm{p}, a_\mathrm{p}$ and $n_\mathrm{p}$ are the planet's mass, orbital radius and orbital frequency, respectively. The gas drag force per unit mass acting on the particle is taken to be that appropriate for high Reynolds number flow (cf. Landau & Lifshitz 1976, Adachi et al. 1976):

$$\varepsilon \mathbf{F} = -K u \mathbf{u}, \qquad \mathbf{u} = \mathbf{v} - \mathbf{v}_\mathrm{gas}, \qquad (16)$$

where $\mathbf{v}_\mathrm{gas}$ is the velocity of the gas which is assumed to be in Keplerian rotation, but partially supported by internal pressure:

$$\mathbf{v}_\mathrm{gas} \simeq (1 - \eta) \mathbf{v}_\mathrm{kep}(r). \qquad (17)$$

Here $\mathbf{v}_\mathrm{kep}(r)$ is the local Keplerian velocity, $v_\mathrm{kep} = \sqrt{GM_\odot/r}$, and $\eta$ is a measure of the radial pressure gradient in the gas. In models of the Solar Nebula, $\eta$ is a few times $10^{-3}$ (cf. Weidenschilling & Cuzzi 1993). The drag coefficient $K$ is proportional to the density of the gas and to the surface-area-to-mass ratio of the particle; it has dimensions of inverse length. It is worth pointing out that the ratio of the drag force to the central force is given by

$$\varepsilon = rK \left(\frac{u}{v_\mathrm{kep}}\right)^2. \qquad (18)$$

For a circular orbit, $u/v_\mathrm{kep} \sim \eta$; for orbital eccentricity $e$ exceeding $\eta$, $u/v_\mathrm{kep} \sim e$. Therefore, in most cases of interest, $(u/v_\mathrm{kep})^2 \ll 1$, so that even fairly large values of $K$ represent a small perturbation to the underlying Keplerian orbit.

For $K = 0$, there exists a constant of the motion, the Jacobi integral,

$$C = -H + n_\mathrm{p}(\mathbf{r} \times \mathbf{v})_z. \qquad (19)$$

The value of $C$ determines the global stability properties of the test particle orbit. It is therefore a natural parameter – analogous to the total energy and angular momentum for the general N-body problem – for monitoring the error of a numerical solution. For $K \neq 0$, $C$ is not constant; nevertheless, we can define a value of this parameter at any instant of time during the evolution of test particle. In the numerical test described below, we quote the errors in $C$ and in the mean longitude of the test particle.

We choose units such that $G = M_\odot + m_\mathrm{p} = a_\mathrm{p} = 1$. In these units, the planet's orbital frequency is unity and its orbital period is $2\pi$. In the numerical tests, the initial conditions of the test particle were $a = 1.5$, $e = 0.1$, $\omega = \pi/2$, $\lambda = 0$ ($\omega$ and $\lambda$ being the longitude of perihelion and the mean longitude, respectively). We used $m_\mathrm{p} = 10^{-4}$, and started the planet at conjunction with the test particle at zero longitude. The gas drag parameter $\eta$ was fixed at 0.005. Integrations were done for different values of $K$ from 0.0 to 0.1. To determine the errors in an individual orbit integration, we ran a second integration using a smaller step size. The smallest step-size, $\tau$, used was 1/1600 of the planet's orbital period, $P$. The errors displayed in Figures 1–3 compare the numerical solution using the indicated step size with a numerical solution obtained with $\tau = P/1600$.

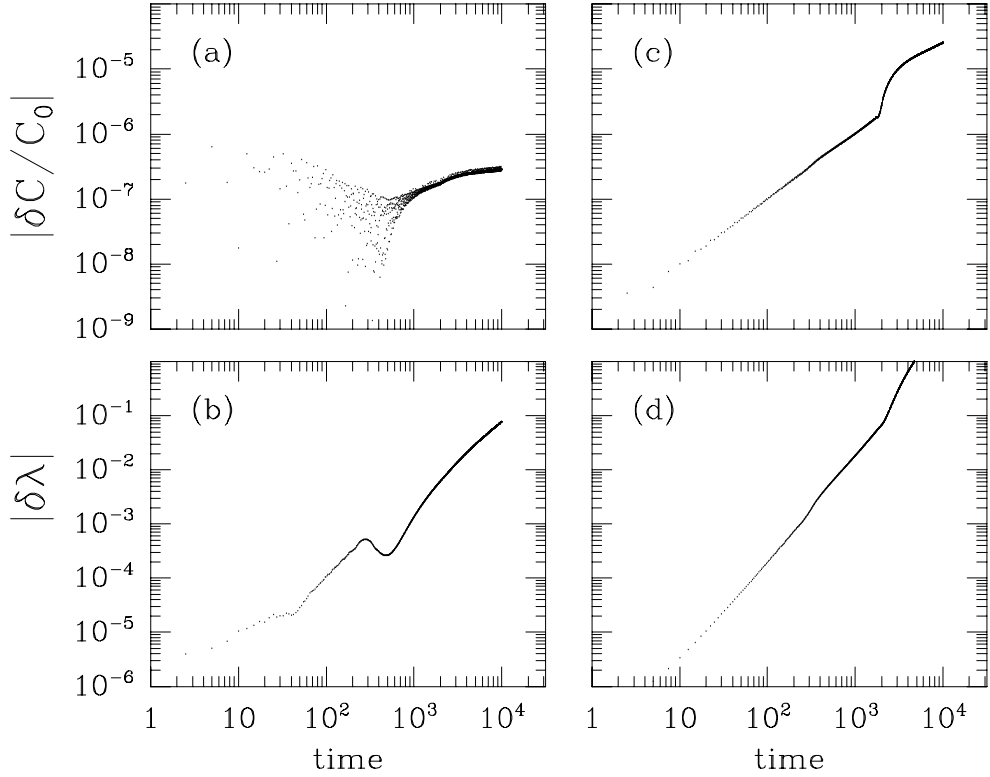

Fig. 1. A test particle orbit subject to a planet's gravitational perturbations as well as gas drag forces is integrated for 10,000$P$, where $P$ is the orbital period of the planet. (The initial conditions and other parameters adopted in this integration are given in the text.) The absolute value of the errors in the instantaneous Jacobi integral, $C$, and in the instantaneous mean longitude, $\lambda$, are plotted against time (indicated in units of $P$). (a) and (b) show the errors incurred with the generalized 'mixed variable mapping' described in the text; (c) and (d) show the errors incurred with a fourth order Runge-Kutta integrator. These integrations were performed with a fixed step size $\tau = P/100$.

In Figures 1 and 2 we display a comparison of the 'mapping' with a fourth order Runge-Kutta (RK4) integrator.[1] The panels (*a*) and (*b*) on the left are for the 'mapping', and the panels (c) and (d) on the right are for the Runge-Kutta integrator. In these integrations we set the drag coefficient $K = 0.01$.

---

[1] We note that RK4 makes four force evaluations per time step, whereas the 'mapping' makes, formally, only one force evaluation per step (and this does not include the central force). However, moving the particle with the $\tilde{f}$, $\tilde{g}$ and $\tilde{h}$ functions consumes slightly more machine time than one force evaluation in RK4. (This estimate can be expected to be somewhat sensitive to the form of the dissipative force.) We find that one time step with the 'mapping' uses approximately half the

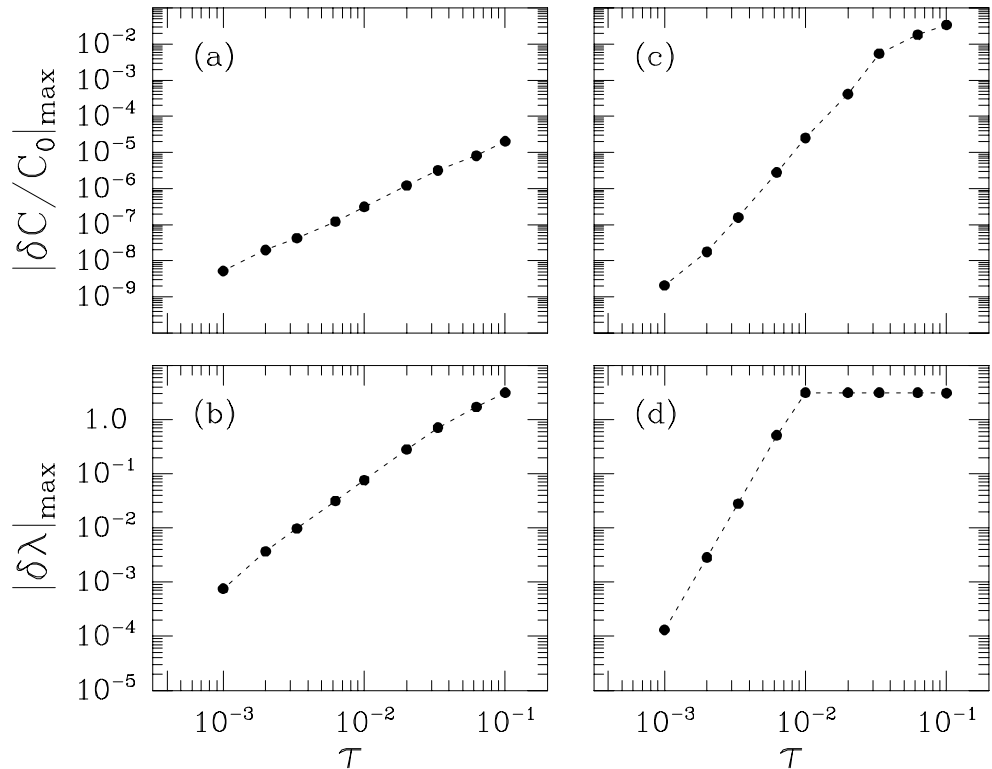

Fig. 2. The maximum error in the Jacobi integral, $C$, and in the mean longitude, $\lambda$, after 10,000 planet orbits when the orbit of Fig. 1 is integrated with different step sizes, $\tau$. (a) and (b) show the errors incurred with the 'mapping'; (c) and (d) show the errors incurred with a fourth order Runge-Kutta integrator.

In Figure 1 we plot the error in $C$ and in the mean longitude, $\lambda$, against time for a run using step size $\tau = P/100$. We observe that the mean longitude error grows linearly with time in the 'mapping' method, but quadratically with time in the RK4 integration. Furthermore, with the 'mapping', there is virtually no secular growth of the error in $C$ (recall that this is also characteristic of the mixed-variable symplectic integrators for the conservative system); with the RK4 integrator, the error in $C$ grows linearly with time. This shows that spurious "numerical dissipation" is virtually absent in the 'mapping'.

Figure 2 shows the maximum errors incurred as a function of step size. We observe that with the 'mapping', the position error increases as $\sim \tau^2$ as does the error in $C$. This shows that the 'mapping' method is of second order. Note also computer time as one step of the RK4. Thus, for the same step size, the 'mapping' is approximately twice as fast as RK4.

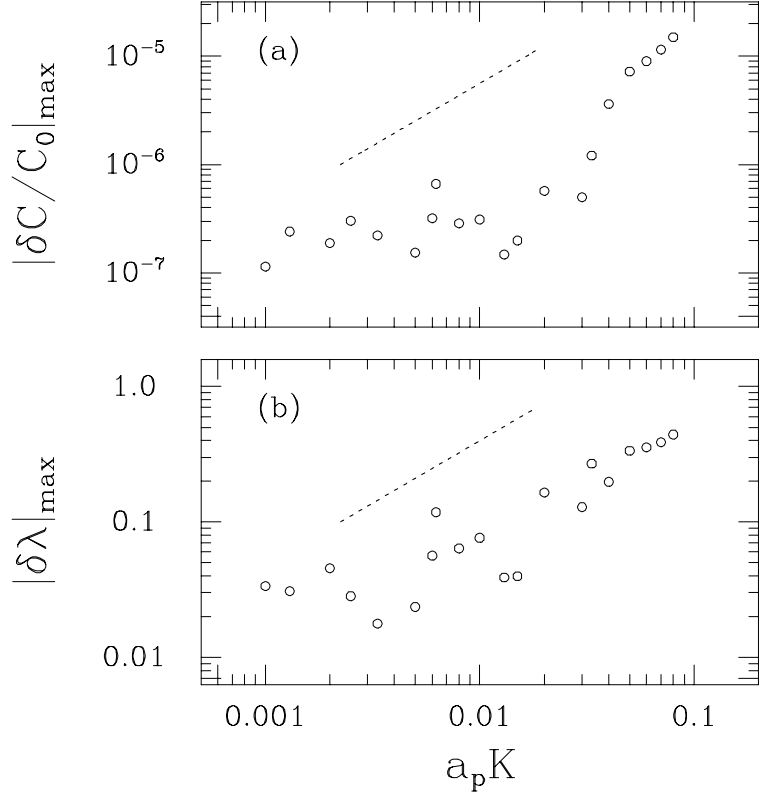

Fig. 3. The maximum error in the Jacobi integral, $C$, and in the mean longitude, $\lambda$, after an integration for 10,000$P$ are shown as a function of the gas drag parameter, $K$. These integrations were done with the 'mapping' with step size $\tau = P/100$. For reference, a linear relationship between the abcissa and the ordinate is indicated by the dashed lines.

that $|\delta\lambda|_{\max}$ and $|\delta C|_{\max}$ both increase as $\sim \tau^{4.5}$ for RK4.[2] However, the numerical factor is dramatically larger than for the 'mapping'. Consequently, the 'mapping' method easily out-performs RK4 for larger step sizes.

In Figure 3 we show the maximum integration error as a function of the magnitude of the drag coefficient, $K$. There is considerable scatter in this plot, but the trend is a roughly linear increase of the error with $K$. This is as expected because in our integrations, the $\mathcal{O}(\varepsilon\tau^3)$ positional error from the drag force is generally comparable to or exceeds the $\mathcal{O}(\mu\tau^3)$ error from the planetary perturbation. (In our numerical tests $\mu \sim 10^{-4}$; the value of $\varepsilon$ (cf. Eqn. 18) varies considerably during each run, but its maximum value is approximately $0.01K$.)

---

[2] It is surprising to find an exponent 4.5, rather than 4 for a fourth-order integrator. We do not have an explanation for this better-than-expected performance of RK4, but see the discussion in Press *et al.* 1989.

## 4. SUMMARY

In this paper, we have introduced a new numerical integration method for Solar system problems with small dissipation. This method is a modification of the second-order "N-body map" introduced recently by Wisdom & Holman (1991) (also independently by Kinoshita et al. (1991)) for nearly-integrable Hamiltonian systems of the form $H = H_{\text{kepler}} + \mu H'$ where $H_{\text{kepler}}$ and $\mu H'$ are separately integrable. $H_{\text{kepler}}$ describes the unperturbed Keplerian motion and $\mu H'$ describes the small mutual gravitational perturbations of the planets (or satellites, in a planetary satellite system).

The new method is summarized as follows. Recall that for the conservative problem, a single step of the N-body map with size $\tau$ requires three operations. These are: (i) evolve the system according to $H_{\text{kepler}}$ for time $\frac{1}{2}\tau$; (ii) evolve the system according to $\mu H'$ for time $\tau$; (iii) evolve the system according to $H_{\text{kepler}}$ again for time $\frac{1}{2}\tau$. The evolution of the unperturbed system, according to $H_{\text{kepler}}$, is efficiently accomplished with the use of Gauss' $f$ and $g$ functions to update the position and velocity of a particle on a Keplerian ellipse. Because the mutual perturbations described by $\mu H'$ are a function of positions alone, sub-step (ii) involves only small increments ("kicks") in the velocities. In the presence of a small dissipative force, $\Delta \ddot{\mathbf{r}} = \varepsilon \mathbf{F}$, we have shown how this integration scheme can be used with *modified* functions $\tilde{f}$ and $\tilde{g}$, and a new function, $\tilde{h}$, to update the positions and velocities in sub-steps (i) and (iii). The corrections, $\tilde{f} - f$, $\tilde{g} - g$ and $\tilde{h}$ are given in terms of the drag force, $\varepsilon \mathbf{F}$.

The method we have described here is a second order integrator: the leading truncation error in the positions per time step $\tau$ is $\mathcal{O}(\mu\tau^3) + \mathcal{O}(\varepsilon\tau^3)$. We have verified this numerically by applying this new method to the restricted three-body problem with a small gas drag force on the test particle. We have given a comparison of the numerical errors of this method with the fourth order Runge-Kutta integrator. The numerical tests show that the position errors grow linearly with the length of integration (in contrast with the quadratic rate of error growth with the Runge-Kutta integrator). Furthermore, the spurious "numerical dissipation" that is present in conventional integrators such as the Runge-Kutta is nearly absent in the 'mapping', as shown by the error in the Jacobi parameter. From the results of numerical tests, we estimate that the 'mapping' is at least an order of magnitude faster than the fourth-order Runge-Kutta for similar levels of integration error.

An obvious disadvantage of the "generalized mixed variable mapping" described in this paper is its low order. For conservative systems, symplectic integrators of higher order are described in Yoshida (1990), Wisdom & Holman (1991) and Kinoshita et al. (1991), and other refinements that yield significantly improved performance are described in Saha & Tremaine (1992). The feasibility of similar improvements in the generalized mapping for dissipative systems will be considered in the future.

Finally, we caution the reader that in planetary dynamics problems with dissipa-tion the evolution of a system is, in general, extremely sensitive to initial conditions and to small changes in system parameters. This sensitivity is an intrinsic physical property of the system, and not merely of numerical solutions. Sensitivity to initial conditions – *aka* chaos – may be found in the conservative system also, but not necessarily in all regions of phase space. In contrast, in the presence of weak dissipation, a nearly-Hamiltonian system is forced to sweep across large regions of phase space, including chaotic regions near resonances. Consequently, sensitivity of the evolution is *always* present in this case, for any initial conditions. The scatter in Figure 3 is indirect evidence of this. It is related to the complicated dynamics of resonance passage phenomena. A discussion of this phenomenon is beyond the scope of this paper. However, it is worth pointing out that resonance passage can produce "anomalous" behavior, *i.e.* evolution that is characterized by singularly large changes of orbital parameters in relatively short times. This phenomenon is well-known in the literature and is an active area of current research (see, for example, Malhotra 1994, for a recent review). The meaning of numerical solutions in the presence of such dynamical behavior is also an interesting question (see, for example, Quinlan & Tremaine 1992).

## Acknowledgements

I thank Jack Lissauer and Martin Duncan for pertinent comments. I am grateful to the organizers of the Planet Formation program at the Institute for Theoretical Physics, Santa Barbara, CA, for their hospitality during the early stages of this work. This research was supported in part by the National Science Foundation under Grant No. PHY89-04035, and was done while the author was a Staff Scientist at the Lunar and Planetary Institute which is operated by the Universities Space Research Association under contract no. NASW-4574 with the National Aeronautics and Space Administration. This paper is Lunar and Planetary Institute Contribution no. 839.